\documentstyle[12pt,moriond]{article}
\begin{document}
\heading{CHEMICAL EVOLUTION OF IRREGULAR GALAXIES}

\vspace{5mm}
\begin{center}
{\normalsize \rm Monica Tosi} \\
{\normalsize \it Osservatorio Astronomico, Bologna, Italy.}\\
\vspace{0.65cm}
\framebox[3.8truecm]
{\rule[-1.9cm]{0.cm}{3.8truecm}}
\vspace{0.35cm}
\end{center}

\begin{moriondabstract}
The state-of-the-art of chemical evolution models for Irregular and Blue
Compact galaxies is reviewed. The resulting scenarios for the initial mass 
function, the regime of star formation and the efficiency of gas outflows
are described: The IMF appears to be roughly the same everywhere; the SF has
possibly followed a bursting regime in BCGs and a more continuous regime in 
Irregulars, but with significant exceptions;
galactic winds must occur in some of these galaxies, with efficiencies
inversely proportional to the galaxy potential well.

\end{moriondabstract}

\section{Introduction}

Irregular (Irr) and Blue Compact (BCG) galaxies have always had significant
importance for cosmological studies. One of the major reasons for this is
that they host the HII regions from which
the relation between helium and oxygen, $\Delta$Y/$\Delta$O,
is derived and, consequently, the primordial helium abundance Y$_p$. They
thus provide a fundamental test for theories of Big Bang Nucleosynthesis.
Moreover, these galaxies are usually poorly evolved systems and can thus
be considered as fairly similar to primordial galaxies. What brings them 
more into fashion nowadays is, however, the circumstance that they have 
been recently suggested as major contributors to the excess of faint blue
objects in deep galaxy counts.

To study properly all the above cosmological issues it would be necessary to 
know not only the current or recent evolutionary conditions of these galaxies, 
but also (or mostly) their conditions at the earlier epochs, which correspond 
to high redshifts. In the case of the Magellanic Clouds and the few other 
closest irregulars in the Local Group, this is feasible with sufficient 
accuracy, thanks to the fact that their old stars are observable and 
examinable in detail. For the great majority of Irrs and BCGs, however, the 
older stars are too faint to be individually resolvable, and the available 
observational data 
refer only to young objects. For these galaxies one of the few ways 
to infer their past evolutionary conditions is from chemical evolution models. 
By selecting those models which are able to reproduce the observed features,
one can in fact derive the most plausible scenarios for previous epochs,
although with the usual caveat \cite {T88} that such scenarios may not be
unique, since different models can provide predictions for the recent times
in similar agreement with the available constraints, but fairly different
predictions for the earlier unconstrained phases \cite {T96}.

\section{Outline of chemical evolution models}

The major task and difficulty for chemical evolution models of Irrs and
BCGs is to reproduce their low observed chemical abundances despite the
relatively high observed star formation rates (SFR).

The theoretical parameters involved in the computation of these models are 
essentially three: the law of star formation (SF), the initial mass function
(IMF), and the gas flows, in and out of the system. Stellar evolution and
nucleosynthesis are the other important ingredients in chemical evolution
models. In the following I concentrate only on the {\it galactic} parameters,
without discussing the uncertainties on stellar evolution theories, despite
the fact that the latter deserve a careful treatment as well.

A priori, the galactic parameters are free 
as a consequence of our ignorance on the 
actual physical mechanisms regulating the formation of stars and the evolution 
of the interstellar medium. However, the comparison with appropriate
observational constraints of the model predictions based on different 
assumptions on the parameters leads to a significant reduction of their
ranges of acceptable values.

The observational constraints usually available for most Irrs and BGCs
are the gas and total masses, $M_g$ and $M_t$ respectively, and the chemical
abundances of the elements measurable from the emission lines of HII regions,
namely: He, N, O and, in a few cases, C and Fe.

Effective constraints on the SF in a galaxy come from its observed
$M_g$/$M_t$ and absolute element abundances. For instance, if a model assumes 
an excessively
high or long SF activity in a given galaxy, this provides an excessively
high number of formed stars, which, in turn, lead to an excessively high gas
consumption and metal production. Such model thus predicts too many
metals and too little gas with respect to the observed values, and this
is the sign that the SF activity must be lower than adopted.

The most significant tests on the IMF are the abundance ratios of elements
produced and ejected by stars of different mass. Fortunately, the few
elements measurable in these galaxies are indeed produced in different stellar
sites. For instance, N is produced mostly by intermediate mass stars, O only
by massive stars, and Fe partly ($\sim$1/3) by SNeII and mostly ($\sim$2/3) 
by SNeIa. Let's assume that for a given galaxy, one adopts a too flat IMF:
there will be too many massive stars with respect to lower mass ones and,
consequently, overproduction of oxygen and underproduction of iron and even 
less nitrogen. The predicted N/O ratio will then be significantly lower and 
the predicted O/Fe higher than observed, a circumstance indicating that the IMF
appropriate for that galaxy should be steeper than the adopted one.

Irrs and BCGs can experience both infall and outflows of gas, depending
on their potential well (generally low) and on the intergalactic medium
where they are imbedded. However, from the point of view of their chemical
evolution, outflows have turned out to be more important than inflows, at
variance with what happens in big spiral galaxies. 
The occurrence of galactic winds
in some Irrs and BCGs was first suggested by \cite{MC} and then confirmed
by several other modellers (see Table 1) as a sort of {\it deus ex machina} able
to overcome the problem that the models overpredicted the metal abundances
of these systems for whatever combination of the other free parameters.
The advantage of the winds is that they can remove the elements in excess,
without significantly altering the other quantities. This obviously does not
imply that galactic winds actually occur, but the comparison of the
observed quantities with the predictions of models with various wind
assumptions can at least tell how the outflows should be.

\section{Description of specific models}

Some years ago, we (\cite{MT} and \cite{MMT}) computed a series of numerical
models with the combined goals of reproducing the observed features of a 
large sample of Irrs and BCGs and of testing the range of acceptable values 
for the theoretical parameters. These
models assumed a bursting mode of SF (i.e. short episodes of intense
activity, separated by long quiescent phases), adopted several IMFs (with
varying slope and/or upper and lower mass cutoffs), and allowed for two
types of gas outflows triggered by SNeII explosions: differential winds 
(i.e. able to remove from the galaxy only the elements synthesized and
ejected by the SNeII) or well mixed winds (i.e. able to remove from the
galaxy a fraction of all the elements present in the interstellar medium 
around the SNe).

\begin{figure}
\vspace{6.5truecm}
\includegraphics{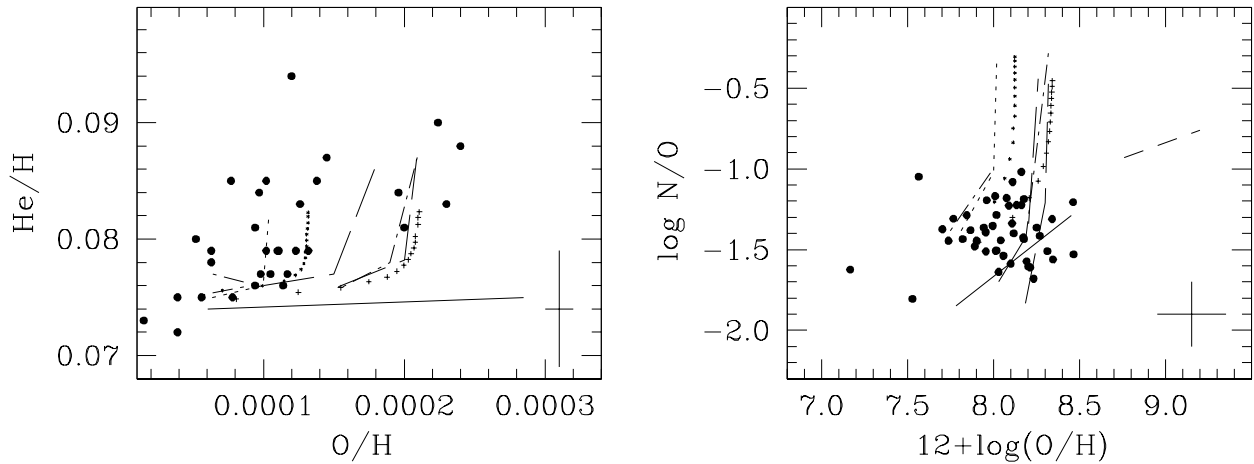}
\caption{Left hand panel: Distribution of the helium vs oxygen abundances
by number as derived from Pagel et al's (1992, \cite{P92}) observations 
of HII regions in Irrs and BCGs 
(dots) and from the predictions of Marconi et al's (1994, \cite{MMT}) models. 
Right hand panel: Distribution of the nitrogen/oxygen vs oxygen abundances
by number for the same data and models as on the left panel.
The average observational error bar is shown in the bottom right corner of
each panel.
}
\label{fig-1}
\end{figure}

Figure 1 shows the predictions of Marconi et al.'s  (1994, \cite{MMT}) models 
for the He vs O and the
N/O vs O abundance ratios compared with the corresponding values derived
from observational data. Models adopting no galactic wind at all are
represented by the short-dashed line: even with only two bursts of SF, oxygen 
is so overproduced that the line falls well outside the observational ranges of 
both distributions (and out of the graph in the left panel) and 
there is no way to bring it within the empirical range of values by varying
SF and/or IMF. This is why in 1985 we (\cite{MT}) introduced winds in our
models. If a well mixed outflow ejecting all the elements  in the same
proportions is assumed (solid line in Figure 1), the predicted oxygen
abundances are lower and consistent with the data, but He and N are reduced 
as well. It can be seen from the two panels that for nitrogen this leads to 
a satisfactory agreement with the data, but helium turns out to be always
underabundant and below the observational range, whatever the assumptions on 
the other free parameters.
If, instead, a differential wind is assumed, ejecting O, but not He and N
that are not produced by SNeII, we can properly reduce the O abundance
without affecting too much either He or N, and predict abundances
consistent with the corresponding data. By varying the wind efficiency
as well as the number and/or the intensity of SF episodes, the models can
span the whole observational ranges of  both distributions.

By running a large number of model cases, \cite{MT} and \cite{MMT} reached
the following conclusions: 
a) each galaxy has a SF and wind history different from
the others;
b) for most Irrs and BCGs, the number of SF
bursts does not exceed 7--10, but some can have sustained 1 or 2 bursts
at most (e.g. IZw18); 
c) the SF bursts have short duration ($\sim 10^8yr$),
whereas the inter-burst phases last several Gyrs; 
d) the IMF is generally universal and around Salpeter's slope.
Similar results have been recently obtained by Bradamante et al. (1998, 
\cite{BMD}) with an updated version of the same code and are described in
this volume.

It is worth emphasizing that all the computed models predict a 
$\Delta$Y/$\Delta$O within each individual galaxy far from being linear.
Qualitatively, there is first a shallow $\Delta$Y/$\Delta$O slope when 
all the O produced in a 
burst is ejected together with the amount of He produced by massive stars; 
but when the lower mass stars, which are He but not O producers, start to die 
and to contribute to the enrichment, He goes up very quickly while O remains 
stable. Quantitatively, this trend is affected by the
adopted efficiency of the winds and of the SF, but is the natural result of
the different lifetimes of the element producers in the context of 
discontinuous star formation and is therefore typical of all dwarfs.
Nonetheless, since the observational data on He and O refer to many galaxies 
with different evolutions, what can be safely derived is only
the average $\Delta$Y/$\Delta$O of the whole galaxy sample. In this situation,
I believe that a linear fit to the observed values remains the best 
statistical approach to infer  the average empirical $\Delta$Y/$\Delta$O.

In 1993, Pilyugin (\cite{P1}) reached very similar conclusions on the 
evolutionary conditions of these galaxies with 
completely independent models. He assumed strongly discontinuous SF,
allowed for the existence of galactic winds, and considered also 
the effect of the HII region self-enrichment. This phenomenon is due to
the fact that when the SNe of a SF burst start to explode, they pollute 
only the gas of their own HII region and not all the surrounding medium;
thus the HII region abundance increases while that of the interstellar medium
remains unaltered. Only after a while, when the HII region mixes with the 
external gas and enriches it, its abundance is diluted and decreases, while
that of the interstellar medium goes up. As a result of this phenomenon, 
Pilyugin predicts 
that the evolution of element abundances in HII regions has a saw-tooth
shape, that appears also in the behaviour of the element abundance ratios.

By comparing his model predictions with the observed distributions of He
vs O and N/O vs O, Pilyugin (\cite{P1}) found that, taken alone, the 
phenomenon of HII region self-enrichment is sufficient to reproduce N and O, 
but not He, whose predicted values as a function of N and O turned out 
always much lower than observed.
Only by introducing differential galactic winds, was he finally
able to reproduce the observed distributions of all the three elements.
He also concluded that Salpeter's IMF is appropriate for Irrs and BCGs.

Carigi et al. (1995, \cite{C}) computed somewhat different models for a sample
of Irrs and BCGs, since they assumed that the SF in these systems is continuous
and follows a law similar to that normally attributed to the solar 
neighbourhood. They also
adopted differential galactic winds and tested various IMFs. As major
constraints for their models they considered three observational quantities:
the $\Delta$Y/$\Delta$O relation, the C/O ratio and the (Z-C-O)/O ratio
(where Z is the total metallicity). Carbon is a useful element to
model, since it is produced mainly by intermediate mass stars, but also
by massive stars. Before the advent of the Hubble Space Telescope no
reliable measures of C in extragalactic HII regions were available and Carigi
et al. were the first who could use such data.
Carigi et al. found that the only models able to simultaneously reproduce
all these constraints are those assuming an age of the galaxies of at least 10
Gyr, a high efficiency of the winds and a steep slope of the IMF at the
low mass end. In particular, they suggested that 23\% of the SN ejecta
should be lost forever through the wind and that the slope of the IMF for 
$M\leq0.5M_{\odot}$ is --2.25, i.e. similar to the extrapolation of Salpeter's,
but quite steeper than that attributed to solar neighbourhood stars
(e.g. \cite{KTG}).

\medskip
\begin{center}
{\bf Table 1.} Chemical evolution models.
\end{center}
\small
\begin{center}
\begin{tabular}{|l|l|c|c|c|c|}
\hline
&&&&&\\
year & authors & objects & IMF &  SF mode &  galactic winds \\
&&&&&\\
\hline
1979 & Lequeux et al & Irr+BCG & -- & episodic & -- \\
1983 & Matteucci \& Chiosi & Irr+BCG & variable ? & episodic & present ? \\
1985 & Matteucci \& Tosi & Irr+BCG & no variations & episodic & necessary \\
1993 & Pilyugin & Irr+BCG & no variations & episodic & selective \\
1994 & Marconi et al & Irr+BCG & no variations & episodic & selective \\
1995 & Carigi et al & Irr+BCG & steeper at low m & continuous & selective \\
1995 & Kunth et al & IZw18 & standard & 1--2 short episodes & selective \\
1991 & Gilmore $\&$ Wyse & LMC & no variations & 2 episodes & no \\
1993 & Russel $\&$ Dopita & MC & steeper & cont or episodic ? & no \\
1995 & Tsujimoto et al & MC & steeper if no wind & cont or episodic ? & could be \\
1996 & Pilyugin & LMC & no variations & 2 long episodes & selective \\
1998 & Pagel $\&$ Tautvaisiene & MC & standard & 2 long episodes & non selective \\
\hline
\end{tabular}
\end{center}
\normalsize 

Table 1 provides a chronological overview of the conclusions that can be found
in the literature for the IMF, SF mode and galactic winds in irregulars of
various types. All these results are derived from the comparison of chemical
evolution models predictions with the observed galactic features.

Table 1 shows that most of the results derived by different authors are
in agreement with
each other. It is interesting to notice, however, that what is found from
studies of the two Magellanic Clouds (MC) is somewhat different from what
is found for dwarf Irrs and BCGs. For instance, the episodes
of SF activity in the MC are several Gyr long, whereas those attributed to
BCGs are at least one order of magnitude shorter. Galactic winds appear
absolutely necessary in dwarfs, whereas there is no obvious need in the MC.
Is this difference real or artificial, and due for instance to the fact that 
the MC are known in more details ? And, if real, is it due to the fact that 
BCGs behave differently from Magellanic Irrs and large samples are dominated 
by BCGs, or to the fact that the LMC is much more massive and large than 
average Irrs ?

With these questions in mind, the results from Table 1 can be summarized as 
follows:
\begin{itemize}
\item The IMF is in general the same everywhere, with slope around Salpeter's;
 which means flatter than that in the solar neighbourhood for high mass stars
 and steeper for very low mass stars.
\item Differential galactic winds are necessary but with varying efficiencies,
 inversely proportional to the galactic mass, so that in massive galaxies
 like the LMC they may be absent.
\item The SF is probably discontinuous, but the significance of the 
 discontinuity remains to be assessed, specially in bigger galaxies like the
 LMC.
\end{itemize}

To have a better insight of the issues related to the galactic winds and
to the SF mode, it is worthwhile to examine these aspects in some further 
detail.

\section{Galactic winds}

From the chemical evolution models described in the previous section, it
appears that galactic winds are necessary to explain the evolution of most,
but not all, irregulars. These were model results with no previous
support on either observational or theoretical ground. In the last
few years, however, there has been an increasing evidence in favour of the
wind occurrence both from observations and from theory. 
First of all, the high velocity filaments observed in H$_{\alpha}$, UV and 
X-rays in several galaxies (e.g. NGC1569, NGC1705, NGC4449, IZw18, \cite{W}, 
\cite{M}, \cite{DC}) have all been interpreted in terms of gas escaping from
the parent galaxy.
Moreover, winds are theoretically predicted by hydrodynamical studies of
the SN ejecta in the interstellar medium typical of Irrs and BCGs (e.g. 
\cite{DG}, \cite{MLF}, \cite{DB}).

\begin{figure}
\vspace{15truecm}
\includegraphics{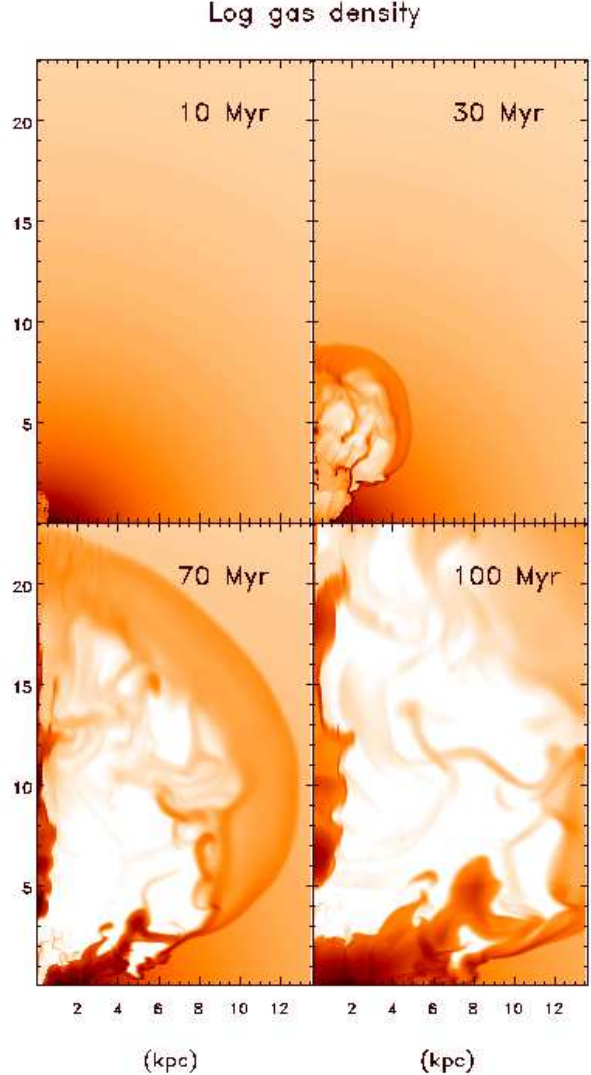}
\caption{Distribution of the gas density in NGC1569 resulting from
SN explosions as predicted by the hydrodynamical models by 
D'Ercole and Brighenti (1998, \cite{DB}). The galaxy major axis x is in 
abscissa, the vertical axis z in ordinate, and both are expressed in kpc.
 See text for details.
}
\label{fig-2}
\end{figure}

D'Ercole and Brighenti (\cite{DB}) are currently performing a detailed
study of the hydrodynamics of the SN ejecta in NGC1569. They have adopted
the literature values for $M_g$=1.1 $\times 10^8$ $M_{\odot}$ and 
$M_t$=3.3$ \times 10^8$ $M_{\odot}$ (\cite{I}) and for the recent 
SFR=0.5 $M_{\odot}yr^{-1}$ (\cite{G}). This SF implies that approximately
30000 SNe have exploded during 30 Myr, ejecting $\sim 10^6$ $M_{\odot}$ of
their own gas. They have then numerically followed the fate of these 
$10^6$ $M_{\odot}$ of gas in the interaction with the interstellar and 
intergalactic media. Figure 2 shows the results of their computations for
the distribution with galactocentric distance of the gas density.  The 
various
grey levels correspond to the gas density in logarithmic scale, in the sense
that the darker, the denser (and white indicates empty regions). The model
predictions are illustrated at four different epochs after the beginning of
the SF activity: 10 Myr in the top left panel, 30 Myr (i.e. 20 Myr after the
previous plot) in the top right panel, 70 Myr in the bottom left panel, and
100 Myr in the bottom right panel.

A few million years after the onset of the SF activity, the SNe start 
to explode, so that at 10 Myr (top left panel) the whitening of the blowing
bubble is already distinguishable toward the galactic center, while the
remaining gas of the galaxy still maintains the original distribution. At 30 
Myr (i.e. at the end of the SF episode) the superbubble has already been able
to go through the galaxy and to expand in the surrounding medium, thus creating
(top right panel) the typical filamentary distribution observed in NGC1569 and 
other Irrs. At 70 Myr, the energy input has stopped since 40 Myr and part of 
the escaped gas is already falling back into the galaxy, as visible from
the dark condensations at the sides of the superbubble in the bottom left 
panel. At 100 Myr the situation has eventually settled down, with some gas
having definitely left the galaxy and the rest having fallen back. From these
models, the estimated amount of gas returning into the galaxy is 
$10^5$ $M_{\odot}$, i.e. only 10\% of that blown out by the SNe; the remaining
90\% is lost forever. Notice that what is lost is only gas from the SN ejecta, 
not from the interstellar medium, thus supporting the differential wind case 
suggested by many chemical evolution models. The total amount of matter 
escaping from NGC1569 is only 1/100 of its $10^8 M_{\odot}$ gas mass.

NGC1569 is a Magellanic irregular with average mass and size, but 
outstanding SF, which implies an unusual energy release in a modest potential 
well. It can therefore be considered as an extreme case of high wind 
efficiency. However, these hydrodynamical results show that even in 
more normal Irrs winds of possibly lower strength may easily occur.

\section{Star Formation Regime}

It was clear right after the first studies of BCGs that their blue colours, 
low metallicities and high gas contents were conceivable only if these galaxies
"are undergoing intermittent and unusually intense bursts of SF" (Searle et 
al. 1973 \cite{SSB}), separated by long quiescent phases. 
The discovery of an underlying old population in most Irrs and BCGs (e.g.
\cite{HG85}) has shown that these galaxies have not started to form stars
only recently, and has therefore reinforced the intermittent SF hypothesis.
The same discovery, however, has also introduced some doubts over this 
scenario, because some of these late type galaxies seem to have formed stars
rather continuously.
The overall scenario that one can currently draw from the literature on 
chemical evolution models and stellar populations studies is that 
irregulars of different type may have different SF regimes. 
Very schematically, the situation could be the following:
\begin{itemize}
\item BCGs undergo bursting SF (e.g. \cite{SSB}, \cite{Hu});
\item dwarf Irrs undergo gasping SF (e.g. \cite{T94}), with long but moderate
 activity separated by short quiescent phases;
\item giant Irrs appear to have continuous SF (e.g. \cite{HG85}), because the 
 quiescent phases tend to be too short to be appreciable.
\end{itemize}

The above scheme can be visualized in the two panels of Figure 3, where the
SF rates as a function of time in BCGs (top left) and Magellanic Irrs (top
right) is sketched. To allow for a more meaningful comparison, the rate of SF 
is normalized per unit area.

This scenario is appealing in what it supports, at least qualitatively, the
hypothesis originally proposed by Gerola et al. (1980, \cite{GS}) with the
theory of stochastic self-propagating SF, in which the bigger the galaxy, the
more continuous the SF. In their theory, in fact, when a {\it cell} starts
forming stars it perturbs its neighbour {\it cells} in such a way that these
start soon to form stars as well. Since a larger galaxy size implies a higher
number of adjacent {\it cells}, the corresponding overall SF inevitably appears
more continuous, because there are higher probabilities that at any time at
least one {\it cell} will be active.

\begin{figure}
\vspace{10truecm}
\includegraphics{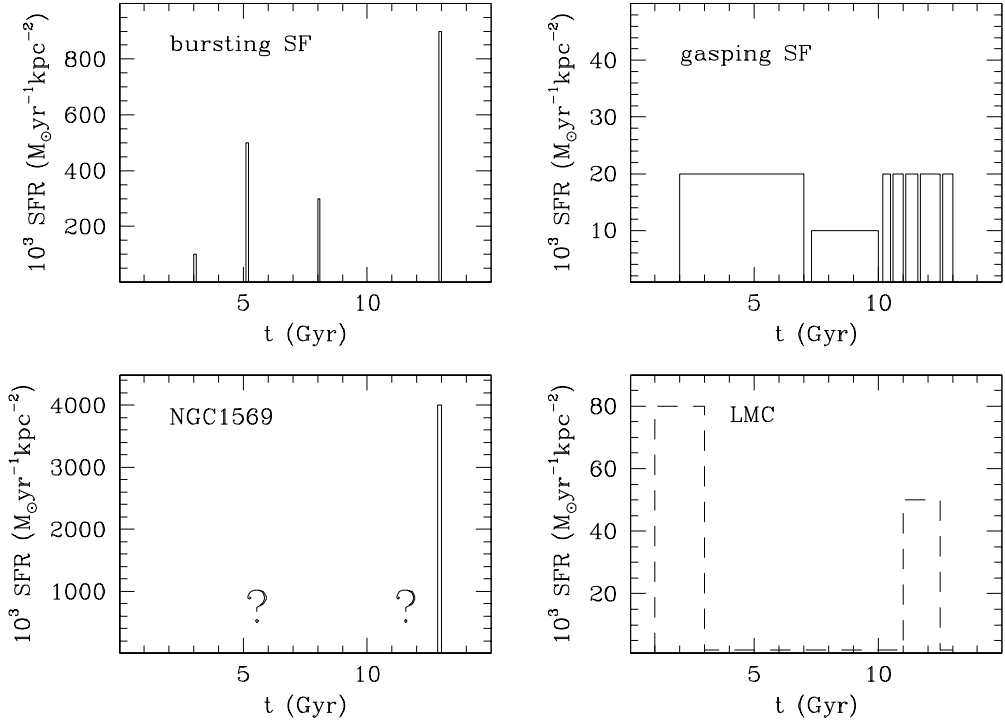}
\caption{Star formation rate per unit area. The top panels sketch the possible
scenarios for BCGs (left) and Magellanic Irrs (right). The bottom panels
show the time distribution of the SF rate as estimated in NGC1569 (left,
\cite{G}) and LMC (right, \cite{L}, \cite{P2}, \cite{PT}). 
}
\label{fig-3}
\end{figure}

Unfortunately, the above scenario is definitely too simplistic, when 
quantitatively compared with the available data on well studied galaxies.
The bottom panels of Fig.3 shows, for instance, an average of the SF rates 
derived by various authors for the LMC (\cite{L}, \cite{P2}, \cite{PT}) 
and NGC1569 (\cite{Hu}, \cite{G}). 

The LMC is one of the largest and most massive irregulars, and has most
probably experienced a SF activity which is not continuous, with two major 
episodes separated by a long period of very low SF rate. The actual rate in the
two episodes and their separation may be uncertain (see also Pagel, this
volume), but all authors favour the indicated scheme. 

NGC1569 is also a
Magellanic irregular, but its small size and mass (e.g. \cite{I}) classify it
as a dwarf Irr. It shows the strongest SF burst ever seen in a normal 
galaxy, with an estimated rate of at least 4 $M_{\odot}yr^{-1}kpc^{-2}$ (\cite 
{G}), one to four orders of magnitude higher than that derived for Local Group 
dwarf Irrs. This high activity makes NGC1569 one of the few local dwarfs 
satisfying the requirements of Babul \& Ferguson's (\cite {BF}) model for
nearby analogs of the faint blue galaxies observed at redshift around
0.5--1. 
The question marks in the plot indicate that some previous
star formation activity should have occurred in NGC1569, as suggested e.g. by 
\cite{VB} from the possible presence of Asymptotic Giant Branch stars, but
it must have been very modest. It is obvious, in fact, from simple gas 
consumption arguments that the recent high activity cannot have taken place 
also in previous epochs, otherwise NGC1569 would have run out of gas long ago, 
whereas 1/3 of its mass is still in gaseous form. It is probable, however,
that much more modest episodes of SF have occurred in the past, to justify
its current metallicity and stellar content. 

We thus face the situation that the prototype of giant irregulars has not
the continuous SF suggested by our hypothetical scenario, and one of the
best studied dwarf Irrs is not gasping at all, but has had what we would have 
rather taken as the paradigma of a BCG burst of SF !

\section {Summarizing the situation ....}

From the above description of the results obtained from chemical evolution
models of irregulars and BCGs, I would conclude that we are now in the
challenging condition of having a qualitative scenario which works acceptably
well for the overall sample of late type galaxies, but is unsatisfactorily
inconsistent with detailed information on important individual systems. 
Fortunately, in these last years we are seeing an incredible improvement in the
quality of the acquired data and on the technical tools available to interpret 
them. I thus think that to improve our understanding of galaxy evolution, rather
than to keep modelling large generic samples of galaxies, we should concentrate
on single representative cases, for which all the necessary information can 
be obtained. In this framework, the suggested flow-chart of future evolution 
models would be the following, for each examined galaxy:

1) get as much information as possible on the evolutionary parameters from
observations; for instance, recent IMF and SF by interpreting spectral and 
photometric data by means of population synthesis models (e.g. \cite{GD})
and synthetic colour-magnitude diagrams (e.g. \cite{G}), gas flows from
optical, UV and/or x-ray data (e.g. \cite{M},\cite{DC}), etc.;

2) compute the hydrodynamics of the interstellar medium and SN ejecta in the
conditions of the examined galaxy;

3) model the chemical evolution of the galaxy by including all the
above data as input.

In this way, the chemical evolution models of each galaxy will be much better
constrained. Once a statistically significant sample of dwarfs will be treated 
in this way, we will be able to draw a really self-consistent scenario for the 
evolution of these systems and to safely infer their average properties at
high redshift.

\vskip 1truecm
{\bf Acknowledgements}: Most of this work results from the collaboration and 
stimulating discussions with Francesca Matteucci. I warmly thank also
Fabrizio Brighenti and Annibale D'Ercole for providing their preliminary
model results and Claus Leitherer for interesting conversations. 
Financial support has been provided by the Italian Space
Agency and by the organizers of this successful meeting.


%
\vfill
\end{document}